\documentstyle[aps]{revtex}

\newcommand{\prepr}[1] {\begin{flushright}  {\bf #1} \end{flushright}
\vskip 1.cm}
\newcommand{\titul}[1] {\begin{center}{\Large {\bf #1 } } \end{center}
\vskip 0.8cm}

\newcommand{\autor}[1] {\begin{center}  {\bf \lineskip .3cm #1  }
                        \end{center} }

\newcommand{\lugar}[1] {\begin{center}  {\normalsize \bf \it #1   }
\end{center}}
\newcounter{muni}

\begin{document}
\hbadness=10000
\pagenumbering{arabic}
\begin{titlepage}
\prepr{NCKU-HEP-00-03 \\
\hspace{20mm} IPAS-HEP-2003 \\
\hspace{20mm} APCTP-00-09 \\
\hspace{20mm} hep-ph/0006yyy}
\titul{\bf Nonleptonic charmless $B$ decays: \\
factorization vs perturbative QCD\footnote{ Some parts of this paper
were presented at the Sapporo Winter School 2000, March 3-7, 2000 in Japan}
}
\autor{Yong-Yeon Keum$^{1}$
\footnote{Email: keum@phys.sinica.edu.tw} and Hsiang-nan Li$^{2}$
\footnote{Email: hnli@mail.ncku.edu.tw}}
\lugar{ $^{1}$ Institute of Physics, Academia Sinica, \\
Taipei, Taiwan 105, Republic of China}
\lugar{ $^{2}$ Department of Physics, National Cheng-Kung University,\\
Tainan, Taiwan 701, Republic of China}

\vspace{10mm}
\begin{abstract}

We compare the factorization approach, the perturbative QCD approach
and the Beneke-Buchalla-Neubert-Sachrajda approach to nonleptonic
charmless $B$ meson decays. We discuss their treatments of factorizable,
nonfactorizable, and annihilation contributions,
infrared-cutoff and scale dependences, and strong phases. 
It is proposed that measurement of CP asymmetries in two-body
$B$ meson decays provides an appropriate test of the these approaches.

\end{abstract}

\vskip 1.0cm
\hspace{15mm}
{\bf  PACS index : 13.25.Hw, 11.10.Hi, 12.38.Bx, 13.25.Ft}

\end{titlepage}
\thispagestyle{empty}

\section{INTRODUCTION}

Exclusive nonleptonic $B$ meson decays provide information of important
dynamics such as penguin contributions, CP asymmetries, and weak and
strong phases. However, it is difficult to analyze these processes
because of their nonperturbative origin. To simplify the analysis, the 
factorization approximation (FA) has been applied to decay amplitudes,
under which nonfactorizable and annihilation contributions are
neglected, and factorizable contributions are expressed as products of
Wilson coefficients, meson decay constants, and hadronic transition
form factors. In this paper we shall compare three approaches to
nonleptonic charmless $B$ meson decays, which are more or less related 
to FA: the FA approach \cite{BSW}, the perturbative QCD (PQCD)
approach \cite{CL,YL,KLS}, and the Beneke-Buchalla-Neubert-Sachrajda
(BBNS) approach \cite{BBNS,Du}. We take the $B\to\pi\pi$ decays as an
example, and explain how factorizable, nonfactorizable, and annihilation 
contributions, infrared-cutoff and scale dependences, and strong phases 
are treated in these approaches.

The effective Hamiltonian for the $B\to\pi\pi$ decays is given by
\cite{REVIEW}
\begin{equation}
H_{\rm eff}={G_F\over\sqrt{2}}
\sum_{q=u,c}V_q\left[C_1(\mu)O_1^{(q)}(\mu)+C_2(\mu)O_2^{(q)}(\mu)+
\sum_{i=3}^{10}C_i(\mu)O_i(\mu)\right]\;,
\label{hbk}
\end{equation}
with the Cabibbo-Kobayashi-Maskawa (CKM) matrix elements
$V_q=V^*_{qd}V_{qb}$ and the four-fermion operators
\begin{eqnarray}
& &O_1^{(q)} = (\bar{d}_iq_j)_{V-A}(\bar{q}_jb_i)_{V-A}\;,\;\;\;\;\;\;\;\;
O_2^{(q)} = (\bar{d}_iq_i)_{V-A}(\bar{q}_jb_j)_{V-A}\;, 
\nonumber \\
& &O_3 =(\bar{d}_ib_i)_{V-A}\sum_{q}(\bar{q}_jq_j)_{V-A}\;,\;\;\;\;
O_4 =(\bar{d}_ib_j)_{V-A}\sum_{q}(\bar{q}_jq_i)_{V-A}\;, 
\nonumber \\
& &O_5 =(\bar{d}_ib_i)_{V-A}\sum_{q}(\bar{q}_jq_j)_{V+A}\;,\;\;\;\;
O_6 =(\bar{d}_ib_j)_{V-A}\sum_{q}(\bar{q}_jq_i)_{V+A}\;, 
\nonumber \\
& &O_7 =\frac{3}{2}(\bar{d}_ib_i)_{V-A}\sum_{q}e_q(\bar{q}_jq_j)_{V+A}\;,
\;\;
O_8 =\frac{3}{2}(\bar{d}_ib_j)_{V-A}\sum_{q}e_q(\bar{q}_jq_i)_{V+A}\;, 
\nonumber \\
& &O_9 =\frac{3}{2}(\bar{d}_ib_i)_{V-A}\sum_{q}e_q(\bar{q}_jq_j)_{V-A}\;,
\;\;
O_{10} =\frac{3}{2}(\bar{d}_ib_j)_{V-A}\sum_{q}e_q(\bar{q}_jq_i)_{V-A}\;, 
\end{eqnarray} 
where $i, \ j$ are color indices and $\mu$ a renormalization scale.

Decay amplitudes are written as
\begin{eqnarray}
{\cal M}&\propto& \langle\pi(P_2)\pi(P_3)|H_{\rm eff}|B(P_1)\rangle\;,
\nonumber\\   
&\propto& \sum_i C_i(\mu)\langle\pi(P_2)\pi(P_3)|O_i(\mu)|B(P_1)\rangle\;,
\end{eqnarray}
with the meson momenta $P_i$. For convenience, we assume
$P_1=(P_1^+,P_1^-,{\bf P}_{1T})=(M_B/\sqrt{2})(1,1,0_T)$,
$P_2=(M_B/\sqrt{2})(1,0,0_T)$, and $P_3=(M_B/\sqrt{2})(0,1,0_T)$, $M_B$
being the $B$ meson mass. The dependences on the renormalization scale
$\mu$ cancel between the Wilson coefficients $C_i(\mu)$ and the hadronic
matrix elements $\langle\pi\pi|O_i(\mu)|B\rangle$, such that the decay
amplitudes are scale-independent (and also scheme-independent). In the
practical derivation of the effective Hamiltonian, an infrared cutoff for
loop integrals needs to be introduced, since four-fermion amplitudes may
contain soft divergences, which are related to nonperturbative dynamics
involved in the matrix elements. If the infrared cutoff appears in the
form that external fermions are off-shell, the derivation will even
become gauge dependent \cite{silver}. These potential infrared-cutoff and
gauge dependences are usually hidden in the matrix elements.

In the FA approach the matrix element of, say, $O_2$ is expressed as
\begin{eqnarray}
\langle\pi(P_2)\pi(P_3)|O_2(\mu)|B(P_1)\rangle&\approx&
\langle\pi(P_2)|(\bar{d}_iq_i)_{V-A}|0\rangle
\langle\pi(P_3)|(\bar{q}_jb_j)_{V-A}|B(P_1)\rangle\;,
\nonumber\\
&\propto& f_\pi F^{B\pi}(q^2=M_{\pi}^2)\;,
\end{eqnarray}
where $f_\pi$ is the pion decay constant and $F^{B\pi}(M_{\pi}^2)$ the
$B\to\pi$ transition form factor evaluated at $q^2=M_{\pi}^2$, $M_\pi$
being the pion mass. The form factor itself is parametrized by models.
Though analyses are simple under FA, a serious problem arises. When decay
amplitudes are expressed as the products of $\mu$-dependent Wilson
coeficients and $\mu$-independent hadronic form factors, they depend on
the scale $\mu$. A cure of this scale dependence has been proposed in
\cite{Ali,Cheng}, which will be discussed in Sec.~II.

The PQCD approach to exclusive $B$ meson decays has been developed some
time ago \cite{L1,LY1}. PQCD is a method to separate hard components from
a QCD process, which are treated by perturbation theory. Nonperturbative
components are organized in the form of hadron wave functions, which can
be extracted from experimental data. This formalism, so-called
factorization theorem, has been successfully applied to various
semileptonic decays \cite{L1,LY1} and nonleptonic (charmed and charmless)
decays \cite{CL,YL,LL,LM,Melic,LUY}. For the $B\to\pi\pi$ decays, the hard
amplitude involves six external on-shell quarks, four of which correspond
to the four-fermion operators and two of which are the spectator quarks 
in the $B$ and $\pi$ mesons. Since nonperturbative dynamics has been
factored out, one can evaluate all possible Feynman diagrams for the
six-quark amplitude straightforwardly, which include both factorizable
and nonfactorizable contributions. Factorizable and nonfactorizable
annihilation diagrams are also included. That is, FA is not necessary in
the PQCD approach. Without FA, the scale independence can be achieved
easily \cite{CLY}. It has been shown that nonperturbative meson wave
functions, being universal, are the same for various topologies of
diagrams in all decay modes, which contain the $B$ and $\pi$ mesons.

Recently, Beneke {\it et al.} proposed an alternative formalism for 
two-body charmless $B$ meson decays \cite{BBNS}. They claimed that
factorizable contributions (the form factor $F^{B\pi}$ in this case) are
not calculable in PQCD, but nonfactorzable contributions are in the heavy
quark limit. Hence, the former are treated in the same way as in the FA
approach, and expressed as products of Wilson coefficients and
$F^{B\pi}$. The latter, calculated as in the PQCD approach, are written
as the convolutions of hard amplitudes with three $(B,\pi,\pi)$ meson
wave functions. Annihilation diagrams are neglected as in FA. Therefore,
this formalism can be regarded as a mixture of the FA and PQCD approaches.
Values of form factors at maximal recoil $q^2=M_\pi^2$ and
nonperturbative meson wave functions are all treated as free
parameters. The BBNS approach then involves more parameters compared to
FA and PQCD. This approach has been applied to the $B\to \pi\pi$ and
$K\pi$ decays \cite{Du}.

In this work we shall make a detailed comparision of the above three
approaches to nonleponic charmless $B$ meson decays. We discuss their
treatments of the infrared-cutoff and scale dependences in Sec.~II,
factorizable contributions in Sec.~III, nonfactorizable contributions in
Sec.~IV, and annihilation contributions and strong phases in Sec.~V. It
will be shown that the CP asymmetry of the $B_d^0\to\pi^\pm\pi^\mp$ decays
predicted in PQCD is much larger than in the FA and BBNS approaches.
Hence, the measurement of CP asymmetries will provide an appropriate
test of these predictions. Section VI is the summary. Note that
final-state-interaction (FSI) effects are always assumed to be absent
in the above three approaches. The neglect of FSI effects has been argued
in \cite{KLS}.

\section{SCALE DEPENDENCE}

As stated in the Introduction, a serious problem of FA is the dependence
of physical predictions on the renormalization scale $\mu$. A plausible
solution to the aforementioned problem has been proposed in
\cite{Ali,Cheng}. The idea is to extract a $\mu$-dependent evolution
factor $g(\mu)$ from the matrix element $\langle\pi\pi|O(\mu)|B\rangle$,
and combine it with the $\mu$-dependent Wilson coefficient $C(\mu)$:
\begin{equation}
{\cal M}\propto C(\mu)\langle\pi\pi|O(\mu)|B\rangle
=C(\mu)g(\mu)\langle\pi\pi|O|B\rangle_{\rm tree}
\equiv C^{\rm eff}\langle\pi\pi|O|B\rangle_{\rm tree}\;,
\label{ude}
\end{equation}
where the effective Wilson coefficient $C^{\rm eff}$ is $\mu$-independent.
The extraction of $g(\mu)$ involves the mixing of effective operators
below the scale $\mu$ and above the lower scale $m_b$, $m_b$ being the
$b$ quark mass. After making a physical amplitude explicitly
scale-independent, FA is applied to the matrix element at the tree level,
\begin{equation}
\langle\pi\pi|O|B\rangle_{\rm tree}\propto f_\pi F^{B\pi}\;.
\label{ude1}
\end{equation}
This strategy has also been adopted in the BBNS approach. 

The above issue is in fact much subtler. One-loop corrections to on-shell
$b$ quark decay amplitudes are usually infrared divergent \cite{silver}.
To remove these infrared divergences, the conventional treatment is to
assume that external quarks are off-shell by $-p^2$. This assumption,
however, results in a gauge dependence of one-loop corrections. The
evolution factor $g(\mu)$ then depends on an infrared cutoff $-p^2$ and
on a gauge parameter, which are originally implicit in the matrix element.
Therefore, the modified FA, though removes the scale dependence of a
decay amplitude, often introduces the infrared-cutoff and gauge
dependences.

The controversy of the scale, infrared-cutoff and gauge dependences can
be resolved in the PQCD approach \cite{CLY}. In this formalism partons,
{\it i.e.}, external quarks, are assumed to be on shell, and both
ultraviolet and infrared divergences in radiative corrections are
isolated using the dimensional regularization. With external quarks being
on shell, gauge invariance of a decay amplitude is guaranteed under
radiative corrections. The ultraviolet poles are subtracted in a
renormalization scheme. The infrared poles, which correspond to
nonperturbative dynamics of decay processes, are absorbed into meson wave
functions. How much finite piece of radiative corrections is absorbed
into wave functions along with the infrared poles depends on a
factorization scheme, or equivalently, on a factorization scale $\mu_f$.
In the PQCD approach $\mu_f$ has been chosen as the inverse of the meson
transverse extent, $1/b$. In other words, dynamics below (above) the
factorization scale is absorbed into a nonperturbative wave function
(perturbative hard amplitude), and a decay amplitude is factorized into
convolution of these two subprocesses. 

For nonleptonic $B$ meson decays, contributions above the factorization
scale involve two scales: $M_W$, at which the matching conditions of the
effective weak Hamiltonian to the full Hamiltonian are defined, and the
typical scale $t$ of a hard amplitude, which reflects specific dynamics
of a decay mode. The scale $t$, corresponding to the lower scale $m_b$
in the modified FA, is chosen as the maximal virtuality of internal
particles in a hard amplitude. The analysis of next-to-leading-order
corrections to the pion form factor \cite{MN} has indicated that this
choice minimizes higher-order corrections.

Without assuming FA, the $\mu$ dependence of a decay amplitude is
calculated based on six-fermion hard amplitudes in the PQCD approach.
Radiative corrections produce various types of logarithms: $\ln(M_W/\mu)$,
$\ln(\mu/t)$ and $\ln(tb)$. The renormalization-group (RG) summation of
the first type of logarithms leads to the Wilson coefficients $C(\mu)$ in
Eq.~(\ref{ude}). The second and third types are summed into the evolution
factor
\begin{equation}
g(\mu)\equiv g_1(\mu,t)g_2(t,1/b)\;.
\label{g12}
\end{equation}
The above expression implies that $g$ in fact contains two pieces, $g_1$
and $g_2$, which are governed by different anomalous dimensions.
The anomalous dimension of $g_1$ is the same as of $C(\mu)$. Below the
hard scale $t$, loop corrections associated with spectator quarks
contribute, such that the anomalous dimension of $g_2$ differs from that
of $g_1$ \cite{CLY}.

Insert Eq.~(\ref{g12}) into Eq.~(\ref{ude}), the effective Wilson
coefficient is written as
\begin{equation}
C^{\rm eff}\equiv C(t)g_2(t,1/b)\;,
\label{eff}
\end{equation}
which consists of two stages of RG evolutions. We have extended the
Wilson evolution directly from $M_W$ down to $t$ without further
considering the mixing of effective operators. The scale independence of
a decay amplitude is thus constructed explicitly. The Wilson evolution is
then followed by the second-stage evolution from $t$ down to $1/b$, which
plays the role of the infrared cutoff $-p^2$.

Together with the second-stage RG evolution, there also exists the
Sudakov evolution $\exp[-s(P,b)]$ from the resummation of double
logarithms $\ln^2(Pb)$ \cite{CS,BS}, which arise from the overlap of
collinear and soft divergences in radiative corrections to meson wave
functions, $P$ denoting the dominant light-cone component of meson
momentum. The effect of the Sudakov evolution will be explained in the
next section. After summing all various large logarithms, the hard
amplitude $H(t)$ can be evaluated at the characteristic scale $t$
perturbatively. $H(t)$ contains all possible factorizable and
nonfactorizable diagrams.

At last, we address the treatment of the dependence on the factorization
scale in the PQCD approach. Both the hard amplitudes and the meson wave 
functions $\phi(x,b)$, $x$ being the momentum fraction associated with a 
valence quark, depend on factorization schemes. For example, $\phi(x,b)$ 
and $\phi(x,b/2)$ are different. However, a decay amplitude, as 
convolution of the two subprocesses, does not depend on factorization 
schemes. The wave functions, though not calculable,
are universal (process-independent), since they absorb long-distance
dynamics, which is insensitive to the specific decay of the $b$ quark
into light quarks with large energy release. In the practical 
application $\phi(x,b/c)$ for the factorization scale $c/b$, $c$ being 
a constant of order unity, can be determined from experimental data of 
some decay modes. These wave functions are then employed to make 
predictions for other modes, whose factorizations are constructed at the 
same scale $c/b$. With this prescription, the dependence on the 
factorization scale is removed. Usually, $c$ is chosen as unity.

A three-scale factorization formula for exclusive nonleptonic $B$ meson
decays is then written as
\begin{eqnarray}
C(t)\otimes H(t)\otimes \phi(x,b)
\otimes g_2(t,b)\otimes \exp\left[-s(P,b)\right]\;.
\label{for}
\end{eqnarray}
Note that Eq.~(\ref{for}) is a convolution relation, with internal parton
kinematic variables $x$ and $b$ integrated out ($t$ depends on $x$ and 
$b$ as defined later). All the convolution factors, except for the wave
functions $\phi(x,b)$, are calculable in perturbation theory. Comparing
Eq.~(\ref{for}) with Eq.~(\ref{ude}), the tree-level matrix element is
read off as
\begin{equation}
\langle\pi\pi|O|B\rangle_{\rm tree}=H(t)\otimes \phi(x,b)
\otimes \exp\left[-s(P,b)\right]\;.
\label{ude2}
\end{equation}
If choosing $t$ as $m_b$, Eq.~(\ref{for}) reduces to a simple product of
the constant Wilson coefficient $C(m_b)$ and the hadronic matrix element.

\section{FACTORIZABLE CONTRIBUTIONS}

In the FA approach factorizable contributions are expressed as products
of Wilson coefficients and transition form factors, which are then
parametrized by some models. This treatment of factorizable contributions
has also been adopted in the BBNS approach. It has been argued that
the $B\to\pi$ transition form factor $F^{B\pi}$ can not be evaluated in
PQCD, because the on-shellness of internal particles in the hard
amplitude is not suppressed by meson wave functions \cite{BBNS}. For
example, the lowest-order hard amplitude in Fig.~1 consists of the
internal $b$ quark and gluon propagators proportional to
\begin{eqnarray}
\frac{1}{[(P_1-k_3)^2-m_b^2](k_1-k_3)^2}&=&\frac{1}{(2P_1\cdot k_3)
(2k_1\cdot k_3)} =\frac{1}{x_1x_3^2 M_B^4}\;,
\label{6} 
\end{eqnarray}
where $k_1$ ($k_3$) is the momentum of the spectator quark in the $B$
($\pi$) meson, and $x_1=k_1^+/P_1^+$ ($x_3=k_3^-/P_3^-$) the
corresponding momentum fraction as indicated in Fig.~1. To derive the
above expression, we have neglected the mass difference
$\bar\Lambda=M_B-m_b$. Obviously, the
pion wave function, if proportional to $x_3(1-x_3)$, does not remove the
power divergence at $x_3\to 0$ from the on-shell internal particles. The
assumption that internal particles in a hard amplitude should be
off-shell is then violated, and PQCD is not applicable to $F^{B\pi}$.

A different viewpoint has been adopted in the PQCD approach. Since the
end-point divergence is not of the pinched type, which is absorbed into
a wave function, we argue that it can be either killed by a pion wave
function vanishing faster than $x_3$ as $x_3\to 0$, or smeared out by
parton transverse momenta $k_T$ \cite{LS}. Including $k_T$, Eq.~(\ref{6})
becomes
\begin{eqnarray}
\frac{1}{(x_3M_B^2+k_{3T}^2)[x_1x_3M_B^2+({\bf k}_{1T}-{\bf k}_{3T})^2]}\;,
\label{6m}
\end{eqnarray}
which has no singularity at $x_3\to 0$. As argued in \cite{LS}, partons
can carry small amount of transverse momenta. When the end-point region
with $x\to 0$ is not important, it is proper to approximate
Eq.~(\ref{6m}) for the hard amplitude by Eq.~(\ref{6}), and the
transverse degrees of freedom of partons in meson wave functions can be
integrated out. When the end-point region is important as in the case of
$F^{B\pi}$, one should keep $k_T$ for a consistent analysis. Once the
transverse degrees of freedom of partons are taken into account, the
factorization of a QCD process should be performed in the $b$ space, $b$
being the variable conjugate to $k_T$ \cite{BS}. This is how the 
factorization scale $1/b$ is introduced in the previous seciton.

As stated before, the scale $t$ should be chosen as the maximal
virtuality of internal particles in a hard amplitude,
\begin{equation}
t=\max(\sqrt{x_1x_3}M_B,\sqrt{x_3}M_B,1/b)\;,
\label{hts}
\end{equation}
according to Eq.~(\ref{6m}). The Sudakov factor $\exp[-s(P,b)]$
suppresses the long-distance contributions in the large $b$ region, and
vanishes as $b=1/\Lambda_{\rm QCD}$, $\Lambda_{\rm QCD}$ being the QCD
scale. This factor then guarantees that most contributions to decay 
amplitudes come from the region with small coupling constant 
$\alpha_s(t)$ \cite{LS}.
It has been shown that 97\% of the contribution to $F^{B\pi}$ arises from
the region with $\alpha_s(t)/\pi <0.3$ \cite{KLS}. It indicates that
$F^{B\pi}$ at large recoil of the pion is well within the perturbative
regime, and that dynamics from hard gluon exchanges dominates. This is
the reason PQCD with Sudakov suppression is applicable to exclusive
decays around the energy scale of the $B$ meson mass \cite{LY1}.
The above conclusion is consistent with the following physical picture.
The $B\to\pi$ transition occurs via the $b\to u$ decay with the outgoing
$u$ quark being energetic. The spectator quark of the $B$ meson is more or
less at rest. The probability that this spectator quark and the $u$ quark
with large relative velocity form the pion is strongly suppressed by the
pion wave function and by the Sudakov factor. Therefore, the process
dominates, where a hard gluon is exchanged so that the two quarks
are aligned to form the pion.

We need to illuminate another concept about the application of PQCD. The
smallness of PQCD predictions compared to experimental data does not
definitely imply that perturbative contribution is not important
\cite{KK}. PQCD predictions depend on wave funcitons one adopts. For
example, a smooth $B$ meson wave function decreases PQCD predictions for
$F^{B\pi}$ by a huge extent. However, this smooth $B$ meson wave function
may not be correct, such that a conclusion drawn from it is not valid. In
our analysis the $B$ meson and pion wave functions are determined from the
data of the pion form factor and of the $B\to D\pi(\rho)$ decays. We then
employ the extracted wave funcitons to predict $F^{B\pi}$ and examine the
dominance of perturbative contributions. In this way the ambiguity in
choosing wave functions and the model dependence of PQCD predictions
are reduced.

Below we highlight the enhancement of penguin contributions observed in
the PQCD approach, and its role in the explanation of the $B\to \pi\pi$
and $B\to K\pi$ data. For simplicity, we present the observation by means
of the FA approach. Consider the ratios $R$ and $R_\pi$,
\begin{eqnarray}
& &R=\frac{{\rm Br}(B^0 \to  K^\pm \pi^\mp)}
{{\rm Br}(B^\pm \to K^0 \pi^\mp)}
=\frac{ a_K^2 + 2 a_K \lambda^2 R_b \cos\phi_3}{a_K^2}\;,
\\
& &R_\pi=\frac{{\rm Br}(B^0 \to  K^\pm \pi^\mp)}
{{\rm Br}(B^0 \to \pi^\pm \pi^\mp)}
=\frac{ a_K^2 + 2 a_K \lambda^2 R_b \cos\phi_3}
{\lambda^2 R_b [R_b +2 a_\pi (R_b -\cos\phi_3)]}\;,
\label{rkpi}
\end{eqnarray}
with the Wolfenstein parameters $\lambda$ and $R_b$, and the Wilson
coefficients
\begin{eqnarray}
a_1&=&C_2+\frac{C_1}{N_c}\;,
\nonumber\\
a_4&=&C_4+\frac{C_3}{N_c}+\frac{3}{2}e_q\left(C_{10}+\frac{C_9}{N_c}
\right)\;,
\nonumber\\
a_6&=&C_6+\frac{C_5}{N_c}+\frac{3}{2}e_q\left(C_8+\frac{C_7}{N_c}
\right)\;,
\end{eqnarray}
$N_c$ being the number of colors. The factor
\begin{eqnarray}
a_{K(\pi)}=\frac{a_4 + 2 r_{K(\pi)} a_6}{a_1}\;,\;\;\;\;   
r_{K(\pi)}=\frac{M_{K(\pi)}^2}{M_B(m_{s(u)}+m_d)}\;,
\label{ptr}
\end{eqnarray}
being negative, represents the ratio of the penguin contribution to the
tree contribution in the $K\pi(\pi\pi)$ mode, where $M_{K(\pi)}$ is the
kaon (pion) mass and $m_q$, $q=s$, $u$, and $d$, the $q$ quark masses.

It is obvious that the data $R\sim 1$ \cite{CLEO3} imply the unitarity
angle $\phi_3 \sim 90^o$, no matter what $a_K$, $\lambda$ and $R_b$ are.
Therefore, $R$ is an appropriate quantity for the determination of
$\phi_3$. While to determine $\phi_3$ from the data of the ratio
$R_\pi\sim 4$ \cite{CLEO3}, one must have precise information of $a_K$,
$a_\pi$, $\lambda$ and $R_b$. It can be shown that the extraction of
$\phi_3$ from $R_\pi$ depends on these parameters sensitively. Hence,
$R_\pi$ is not an appropriate quantity for the determination of $\phi_3$.
To explain $R_\pi\sim 4$, a large $|a_{\pi}|\sim 0.1$ corresponding to
$m_d=2m_u=3$ MeV and a large $\phi_3\sim 130^o$ must be adopted in the FA
approach \cite{WS}. This is obvious from Eq.~(\ref{rkpi}), since a large
$\phi_3$ leads to a constructive interference between the first and
second terms in the numerator of $R_\pi$. In the modified FA approach
with effective number of colors $N_c^{\rm eff}$, a large unitarity angle
$\phi_3\sim 105^o$ is also concluded \cite{HYCheng}.

It is interesting to ask whether one can explain $R_\pi\sim 4$ using
reasonable quark masses $m_d=2m_u\sim 10$ MeV and a smaller
$\phi_3\sim 90^o$, if the extraction from $R$ is given a higher weight.
The answer is positive in the PQCD approach. An essential difference
between the FA and BBNS approaches and the PQCD approach is that
transition form factors can be calculated in the latter as argued above.
Therefore, we do not assume that the form factors associated with the
Wilson coefficients $a_{1,4}$ and with $a_6$ are the same. An explicit
PQCD calculation indicates that the ratios $a_{K,\pi}$ are more
complicate, which reduce to Eq.~(\ref{ptr}) only in some extreme
kinematic conditions of partons \cite{KLS}. Moreover, the choice of the
hard scale $t$, at which Wilson coefficients are evaluated, is also
different. It is chosen arbitrarily as $m_b$ or $m_b/2$ in the FA and
BBNS approaches, but as virtuality of internal particles of a hard
amplitude in PQCD, such that evolution effects vary among the above form
factors.

It has been noticed that $a_1(\mu)$, $a_4(\mu)$ and $a_6(\mu)$ have
dramatically different $\mu$ dependences for $\mu< m_b/2$: $|a_4|$ and
$|a_6|$ increase with the decrease of $\mu$ faster than $a_1$. The choice
of the scale in PQCD makes possible that $t$ runs to below $m_b/2$,
and that penguin contributions are enhanced over tree contributions. It
has been observed that the evolution effect enhances the ratios
$|a_{K,\pi}|$ by about 50\% \cite{KLS}. Hence, even adopting
$m_d=2m_u\sim 10$ MeV, we can account for $R_\pi\sim 4$ using
$\phi_3\sim 90^o$. That is, the data of $R_\pi$ do not demand large
$\phi_3$. We emphasize that the hard scale $t$ with an average 1.4 GeV,
corresponding to $\alpha_s(t)/\pi\sim 0.13$, is still large enough to
guarantee the applicability of PQCD \cite{KLS}.

\section{NONFACTORIZABLE CONTRIBUTIONS}

Nonfactorizable contributions are neglected in the FA approach. However,
it has been known that this approximation is not always appropriate,
especially for the decay modes, whose factorizable amplitudes are
proportional to the Wilson coefficient $a_2=C_1+C_2/N_c$. Since $a_2$ is
small, factorizable contributions and nonfactorizable contributions are
of the same order, and the latter become important. This is the reason it
is difficult to resolve the controversy in the branching ratios of the
$B\to J/\psi K^{(*)}$ decays in FA \cite{YL}. The conventional way to
introduce nonfactorizable contributions is to vary number of colors in
the Wilson coeficients $a_i$. The resultant effective number of colors,
$N_c^{\rm eff}$, being process-dependent parameters, render the FA
approach less predictive.

To go beyond FA, it has been proposed to evaluate nonfactorizable
amplitudes for two-body charmless $B$ meson decays using PQCD
factorization theorem in the BBNS approach \cite{BBNS}. The motivation is
that the power divergence appearing in $F^{B\pi}$ does not exist in
nonfactorizable amplitudes, since soft divergences cancel between the two
diagrams in Fig.~2 \cite{BBNS}. Hence, these diagrams contribute to the
hard parts of the $B\to\pi\pi$ decays, and are calculable in PQCD. We
emphasize that the above soft cancellation is not sufficient to justify
the applicability of PQCD. No matter what value $M_B$ is, one can always
obtain the soft cancellation between the two nonfactorizable diagrams.
However, it is obvious that PQCD is not applicable to $B$ meson decays if
$M_B$ is as low as 1 GeV. The criterion for the applicability of PQCD is
stronger: PQCD is applicable only when higher-order corrections to
decay amplitudes are under control.

We shall show explicitly, following the reasoning in \cite{IL}, that the
simple factorization formula in \cite{BBNS} for a nonfactorizable
amplitude suffers large higher-order corrections. The expression is
\begin{equation}
f=C \alpha_s\int \frac{dx_1}{x_1}\phi_B(x_1)
\left[\int\frac{dx}{x}\phi_\pi(x)\right]^2\;,
\label{1}
\end{equation}
where the constant $C$ is related to a color factor, the denominator
$x_1 x$ comes from the hard gluon propagator, and another denominator $x$
comes from the internal quark propagator. If the hard part is
characterized by $M_B$, the coupling constant $\alpha_s$ evaluated at the
scale $\mu\sim M_B$ is small and higher-order corrections are negligible.
However, it is not the case. When including higher-order
corrections that cause the running of $\alpha_s$, the logarithm
$\alpha_s(\mu)\ln(x_1 x M_B^2/\mu^2)$ occurs, with $x_1 x M_B^2$ being
the invariant mass of the hard gluon. In order to diminish this
logarithmic correction, the appropriate choice of the scale is
$\mu=\sqrt{x_1 x}M_B$. The coupling constant $\alpha_s$ in Eq.~(\ref{1})
becomes running and should move into the convolution relation, leading to
\begin{equation}
f=C \int \frac{dx_1}{x_1}\phi_B(x_1)
\frac{dx_2}{x_2}\phi_\pi(x_2)\frac{dx_3}{x_3}\phi_\pi(x_3)
\alpha_s(\sqrt{x_1 x_3}M_B)\;.
\label{2}
\end{equation}

Assume the following models for the meson wave functions:
\begin{eqnarray}
\phi_B(x)&=&N_B\sqrt{x(1-x)} \exp\left(-\frac{M_B^2 x^2}
{2\omega_B^2}\right)\;,
\label{3}\\
\phi'_B(x)&=&N'_Bx^2(1-x)^2 \exp\left(-\frac{M_B^2 x^2}
{2\omega_B^{\prime 2}}\right)\;,
\label{35}\\
\phi_\pi(x)&=&\frac{\sqrt{6}}{2}f_\pi x(1-x)\;.
\label{4}
\end{eqnarray}
The $B$ meson wave functions $\phi_B$ \cite{BW} and $\phi'_B$ \cite{KLS}
possess peaks at small $x$ that depend on the shape parameters $\omega_B$
and $\omega'_B$, respectively. According to \cite{BBNS}, the values of
the shape parameters can be fixed via the parameter $\lambda_B=0.3$ GeV:
\begin{equation}
\frac{\int_0^1 (dx/x)\phi_B(x)}{\int_0^1 dx\phi_B(x)}=
{M_B \over \lambda_B}=17.6\;,
\end{equation}
which gives $\omega_B=0.65$ GeV and $\omega'_B=0.25$ GeV. The
normalization constant $N_B$ is related to the $B$ meson decay constant
$f_B$ through $\int_0^1\phi_B(x)dx=f_B/(2\sqrt{2N_c})$. The asymptotic
pion wave function in Eq.~(\ref{4}) has been adopted in \cite{BBNS}.

Using the wave function $\phi_B$($\phi'_B$), we find that only about 
20\% (30\%) of the full contribution to the nonfactorizable amplitude 
comes from the perturbative region with $\alpha_s/\pi < 0.3$.
To derive the above percentages, we have chosen 
$\Lambda_{\rm QCD}=250$ MeV for the running $\alpha_s$, and frozen the
scale $\sqrt{x_1 x_3}M_B$ to 0.4 GeV, as it goes to below 0.4 GeV. We
conclude that Eq.~(\ref{2}) is dominated by soft contribution from the
end-point region with $x_1,x\to 0$, {\it i.e.}, with $\alpha_s >1$, where
the perturbative expansion is not justified. Therefore, nonfactorizable
contributions can not be evaluated reliably in the BBNS approach. This
conclusion is consistent with that drawn in \cite{IL}: PQCD without
Sudakov suppression is not applicable to exclusive processes for an
energy scale below 10 GeV. The formalism in \cite{BBNS} is
self-consistent as $M_B\to \infty$, but not in the case with $M_B\sim 5$
GeV.

In the end-point region where soft contributions are important,
nonfactorizable amplitudes should be treated in a more careful way. That
is, the dependence on the parton transverse momentum $k_T$ is essential,
and should be included as argued in the previous section. The
argument of $\alpha_s$ is also set to $t$, the maximal virtuality of
internal particles of a hard amplitude, which is similar to that in
Eq.~(\ref{2}), but depends on the additional scale $1/b$. $\alpha_s(t)$
is then always small even in the end-point region, since $1/b$ is large
under Sudakov suppression \cite{LS}. It has been shown that almost 100\% 
of full contributions to nonfactorizable amplitudes arise from the 
region with $\alpha_s(t)/\pi < 0.3$. That is, the evaluation of 
nonfactorizable contributions is reliable in the PQCD approach.

There is another difference between the BBNS and PQCD approaches in the
treatment of nonfactorizable contributions. The momentum of the light
spectator quark in the $B$ meson has been ignored in the former
\cite{BBNS}, such that quark propagators in the hard part always remain
time-like:
\begin{equation}
\frac{1}{(k_2+k_3-k_1)^2+i\epsilon}\approx \frac{1}{(k_2+k_3)^2}
= \frac{1}{x_2x_3M_B^2}\;.
\end{equation}
This approximation drops the strong phase from the kinematic region
where the virtual quark becomes on-shell, $(k_2+k_3-k_1)^2\sim 0$.
In the PQCD approach nonfactorizable contributions are complex:
\begin{equation}
\frac{1}{(k_2+k_3-k_1)^2+i\epsilon}= 
P \left[ \frac{1}{(k_2+k_3-k_1)^2} \right]
-i\pi\delta \left[(k_2+k_3-k_1)^2\right] \;,
\end{equation}
where $P$ denotes the principle-value prescription. As shown in
\cite{LUY}, the above imaginary part is not the main source of strong
phases for the $B\to\pi\pi$ decays, because nonfactorizable contributions
are only few percents of factorizable ones. However, it will become
important in decay modes whose factorizable amplitudes are proportional
to the small Wilson coefficient $a_2$, or absent, such as the
$B_d^0\to K^\pm K^\mp$ decays \cite{CCL}.

\section{STRONG PHASES}

In the FA and BBNS approaches annihilation contributions have been
neglected. Though annihilation amplitudes from the operator $O_4$ for 
the $B\to \pi\pi$ decays indeed vanish because of helicity suppression, 
those from $O_6$ do not. As shown in Eqs.~(B1) and (B6) of \cite{HYC},
there is a large enhancing factor 
\begin{equation}
\frac{M_B^2}{(m_b+m_d)(m_u+m_d)}\;,
\end{equation}
associated with annihilation contributions from $O_6$, which is much 
larger than the one
\begin{equation}
\frac{M_\pi^2}{(m_b+m_d)(m_u+m_d)}\;,
\end{equation}
associated with usual penguin contributions. Though annihilation
contributions are suppressed by a factor $f_B/M_B$ \cite{Brod},
they can be comparible with penguin contributions. It has been observed
that the annihilation diagrams Figs.~3(e) and 3(f) contribute large
imaginary parts (strong phases) in the PQCD approach \cite{KLS}.

In the approaches where annihilation contributions are neglected, strong
phases must be introduced via the Bander-Silverman-Soni (BSS) mechanism
shown in Fig.~4(a) \cite{BSS}. However, this mechanism, contributing
through a charm quark loop, is suppressed by the charm mass threshold
\cite {KLS}:
\begin{equation}
q^2=(x_2P_2+x_3P_3)^2=x_2x_3 M_B^2 > 4 m_c^2\;,
\end{equation}
$m_c$ being the $c$ quark mass. Obviously, the large $x_2$ and $x_3$
regions are not favored by the pion wave funtion in Eq.~(\ref{4}).
. Moreover, the diagram with a charm quark loop is of higher order in 
the PQCD formalism. An exact numerical analysis has indicated that the
imaginary contribution from the BSS mechanism is smaller than that
from the annihilation diagrams by a factor 10 \cite{KLS}.

Another source of strong phases in the FA and BBNS approaches is the 
diagrams in Fig.~4(b), whose imaginary contributions appear in the 
effective Wilson coefficients in Eqs.(4)-(8) of \cite{BBNS}, for instance, 
\begin{eqnarray}
a_1^{u} &=& C_1 +{C_2 \over N} + {\alpha_s \over 4 \pi}{C_F \over N}
C_2 F\;,
\nonumber\\
F &=& -12 \ln{\mu \over m_b} - 18 + f_{\pi}^{I} + f_{\pi}^{II}\;,
\label{a1f}
\end{eqnarray}
with
\begin{equation}
f_{\pi}^{I} = -0.5 - 3 \pi i\;,
\label{fpi}
\end{equation}
and $f_\pi^{II}$ being real. We emphasize that Eq.~(\ref{fpi}) is derived
from a configuration, in which the outgoing $\bar u$ quark carries the
full pion momentum. Again, this configuration is strongly suppressed by
the pion wave function and by the Sudakov factor. That is, the imaginary
contribution from the above source has been overestimated.

Below we make a quantative comparision among the FA, BBNS and PQCD
approaches. Figure 3 collects all the Feynman diagrams for the hard
amplitude of the $B_d^0 \to \pi^{+}\pi^{-}$ decay, including 
factorizable and nonfactorizable tree, penguin and annihilation diagrams, 
whose contributions are listed in Table I. The results obtained in the 
BBNS approach are also shown. The FA results
are the same as the BBNS ones, but with the nonfactorizable amplitudes
neglected. We observe
\begin{enumerate}
\item
The tree contributions are almost equal in the three approaches, but
the penguin contributions are larger in PQCD, corresponding to a larger
ratio $|a_\pi|$ defined by Eq.~(\ref{ptr}) (about twice of those in FA
and BBNS), {\it i.e.}, corresponding to the dynamical penguin enhancement. 

\item
The tree and penguin contributions are real in PQCD, but complex in
FA and BBNS due to the complex effective Wilson coefficients, which are 
the main source of strong phases in FA and BBNS.

\item
The annihilation contributions are neglected in FA and BBNS, but of the 
same order as the penguin contributions in PQCD, which are the main 
source of strong phases in PQCD.

\item
Nonfactorizable contributions are complex in PQCD but real in BBNS, as 
explained in the previous section, and neglected in FA. Note that 
different $B$ and $\pi$ meson wave functions have been employed as 
deriving the results in Table I \cite{KLS,BBNS}. Nonfactorizable 
contributions are destructive relative to the penguin contributions, if 
concering only the real parts, and negligible for the $B\to\pi\pi$ 
branching ratios in both approaches.

\item
The dominant nonfactorizable contribution $M^P$ in BBNS is due to the 
sharp hard parts in the end-point regions of momentum fractions $x$
as stated in Sec.~III.

\item
The total tree and penguin contributions, including both factorizable,
nonfactorizable and annihilation ones, are
\begin{eqnarray}
& &T_{\rm PQCD}=(71.87+3.20i)\times 10^{-3}\;,\;\;\;\;
P_{\rm PQCD}=(-7.47 + 4.64 i)\times 10^{-3}\;,
\nonumber\\
& &T_{\rm BBNS}=(74.93+1.13i)\times 10^{-3}\;,\;\;\;\;
P_{\rm BBNS}=(1.13 - 1.27 i)\times 10^{-3}\;,
\nonumber\\
& &T_{\rm FA}=(75.70+1.13i)\times 10^{-3}\;,\;\;\;\;
P_{\rm FA}=(-3.04 - 1.27 i)\times 10^{-3}\;.
\label{ppbb}
\end{eqnarray}

\end{enumerate}

The tree contributions obtained in the three approaches are very close.
The signs of the penguin contributions are different, and the magnitude 
in PQCD is larger than those in FA and BBNS:
\begin{equation}
|P_{\rm PQCD}|\sim 5.2 |P_{\rm BBNS}|\;,\;\;\;\;
|P_{\rm PQCD}|\sim 2.7 |P_{\rm FA}|\;.
\label{pra}
\end{equation}
The difference in the total penguin contributions is attributed 
to the treatment of nonfactorizable and annihilation duagrams. We 
emphasize that the BBNS results depend on the parameters for 
factorizable contributions (the transition form factors) and the 
parameters for nonfactorizable contributions (the meson wave funcitons). 
Therefore, it is possible to reverse the sign and to increase the 
magnitude of the penguin contribution by varying these parameters. 
The penguin contribution in the BBNS approach is smallest because of
the cancellation between $F^P$ and $M^P$. 
If expressing the amplitudes of the $B_d^{0} \to \pi^+\pi^-$ decay as
\begin{equation}
{\cal A}= V_{t}^{*} |P| e^{i\delta} -V_{u}^{*} |T|,
\end{equation}
with $P$ and $T$ given in Eq.~(\ref{ppbb}), the strong phases $\delta$ 
are about
\begin{equation}
\delta_{\rm PQCD} \sim 123^o\;,\;\;\;\;
\delta_{\rm BBNS} \sim -50^o\;,\;\;\;\;
\delta_{\rm FA} \sim -113^o \;.
\label{sp}
\end{equation}

The predictions for the CP asymmetry of the $B_d^0\to\pi^\pm\pi^\mp$ 
decays are presented in Fig.~5. Because of Eq.~(\ref{pra}), the CP 
asymmetry in PQCD is larger than that ($\sim 10 \%$ at most) in FA 
and much larger than that ($\sim 4 \%$ at most) in BBNS. 
Because of Eq.~(\ref{sp}), the sign of the CP asymmetry in PQCD 
is opposite to those in FA and BBNS. If the $B$ meson wave function in 
\cite{BW} with $\omega_B=0.4$ GeV is adopted, $M^P$ and also 
$|P_{\rm BBNS}|$ will increase accordingly, such that the CP asymmetry 
in the BBNS approach becomes a bit larger. 
We shall leave the resolution of the above 
discrepancies to future measurements of CP asymmetries. That is, 
experimental data of CP asymmetries will provide an appropriate test of 
the three approaches. Note that all the three approaches give the 
similar branching ratio 
$B(B_d^0\to \pi^\pm \pi^\mp)\sim 4.6\times 10^{-6}$, which is 
consistent with the CLEO data \cite{CLEO3}.

As argued in Sec.~III, to account for the $B\to K\pi$ and $\pi\pi$ data 
simulataneously using the unitarity angle $\phi_3=90^o$, the ratio of 
the penguin contribution to the tree contribution must be as large as 
0.1. It is easy to find 
\begin{equation}
|P_{\rm PQCD}|/|T_{\rm PQCD}|= 0.12\;,\;\;\;\;
|P_{\rm BBNS}|/|T_{\rm BBNS}|= 0.02\;,\;\;\;\;  
|P_{\rm FA}|/|T_{\rm FA}|= 0.04\;,
\end{equation}
from Eq.~(\ref{ppbb}). Due to the 
smaller ratio in the BBNS and FA approaches, a larger angle 
$\phi_3> 90^o$ must be adopted in order to explain the $B\to\pi\pi$
data \cite{Du,WS}. As a consequence of the penguin enhancement, the CP 
asymmetry is also large in the PQCD approach. 

\section{SUMMARY}

We summarize the comparision among the FA, BBNS, and PQCD approaches
as follows:
\begin{enumerate}
\item
The scale independence of predictions in the FA and BBNS approaches
is achieved by the generalized FA method: a $\mu$-dependent evolution
factor is extracted from hadronic matrix elements by considering the
mixing of the effective operators at a scale below $\mu$. This evolution
factor is then combined with the $\mu$-dependent Wilson coefficient
to form a scale-independent effective Wilson coefficient. However, the
infrared-cutoff and gauge dependences appear. In the PQCD approach the
scale independence is made explicit by extending the Wilson evolution
down to the hard scale $t$ directly, which is then followed by the
second-stage RG evolution evaluated based on six-quark amplitudes. Since
external quarks are on-shell, the gauge invariance of PQCD predictions is
guaranteed. The infrared-cutoff dependence, treated as a factorization 
scheme dependence, is removed by the universality of wave functons.

\item
Factorizable contributions (hadronic transition form factors) are not
claculable in the FA and BBNS approaches, but calculable in the PQCD
approach. The inclusion of parton transverse momenta smear the end-point
singularity, and Sudakov suppression guarantees that almost all 
contributions come from the region with $\alpha_s/\pi< 0.3$.

\item
Dynamical penguin enhancement exists in the PQCD approach, but not in 
the FA and BBNS approaches. This enhancement, arising from different
evolution effects in the form factors associated with the Wilson 
coefficient $a_{1}$, $a_{4}$ and $a_{6}$, is cruical for the 
simultaneous explanation of the $B\to K\pi$ and $\pi\pi$ data using a
smaller  unitarity angle $\phi_3\sim 90^o$ \cite{KLS}.

\item
Nonfactorizable amplitudes are neglected in the FA approach, but
calculable in the BBNS and PQCD approaches. However, without Sudakov
suppression, higher-order corrections are not under control in the BBNS
approach. The end-point singularity needs to be smeared by taking into
account parton transverse momenta and Sudakov suppression. On the other 
hand, nonfactorizable amplitudes are real in the BBNS approach, but 
complex in the PQCD approach. The conclusion is the same: nonfactorizable
contributions are destructive relative to the penguin contributions
and negligible for the evaluation of branching ratios.

\item
Annihilation contributions are neglected in the FA and BBNS approaches,
but calculable in the PQCD appraoch. The helicity suppression applies
only to amplitudes associated with $O_4$, but not to those associated 
with $O_6$. It is found that the $O_6$ contribution is of the same
order as the penguin contribution, and  gives a large imaginary part.

\item
Strong phases come from the BSS mechanism and the extraction of
the $\mu$ dependence from hadronic matrix elements in the FA and BBNS
approaches, but from annihilation diagrams in the PQCD approach.
It has been argued that the two sources of strong phases in the
FA and BBNS approaches are in fact strongly suppressed by the
charm mass threshold and by the end-point behavior of meson wave functions.

\item
Because of the large penguin and annihilation contributions in the PQCD
approach, the predicted CP asymmetry in the $B_d^0\to\pi^\pm\pi^\mp$ 
decays dominates over those in the FA and BBNS approaches as shown in 
Fig.~5. Future measurements of CP asymmetries can distinguish the FA, 
BBNS, and PQCD approaches.

\end{enumerate}
\vskip 1.0cm

\centerline{\bf Acknowledgment}
\vskip 0.5cm

We thank M. Bando, Y. Sumino and M. Yamaguchi for organizing the Summer
Institute 99 at Yamanashi, Japan, whose active atmosphere stimulated 
this work, H.Y. Cheng, T. Mannel and A.I. Sanda for helpful discussions, 
and M. Kobayashi for his encouragement. The works of HNL and YYK were 
supported in part by the National Science Council of R.O.C. under the 
Grant Nos. NSC-89-2112-M-006-004 and NSC-89-2811-M-001-0053, respectively, 
and in part by the Monbusho visiting fellowship from the Ministry of
Education, Science and Culture of Japan.
\newpage


\newpage

\vskip 3.0cm
{\bf \Large Figure Captions}

\begin{enumerate}
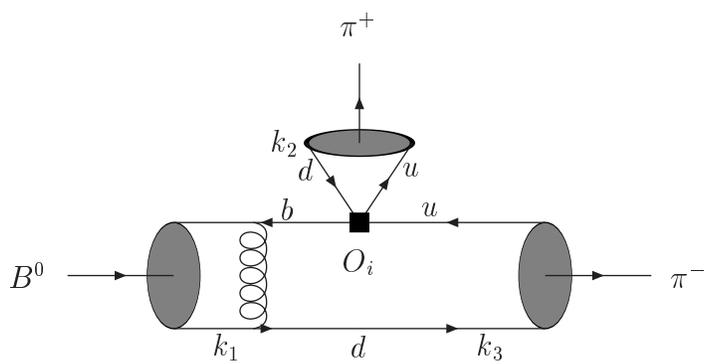
\item Fig. 1: One of the lowest-order factorizable diagrams for
the $B_d^0 \to \pi^{+}\pi^{-}$ decay.

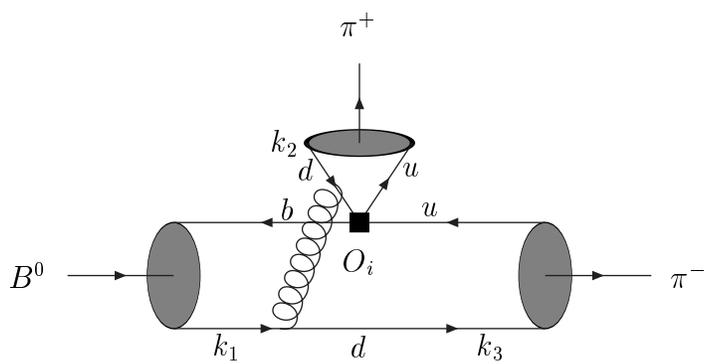
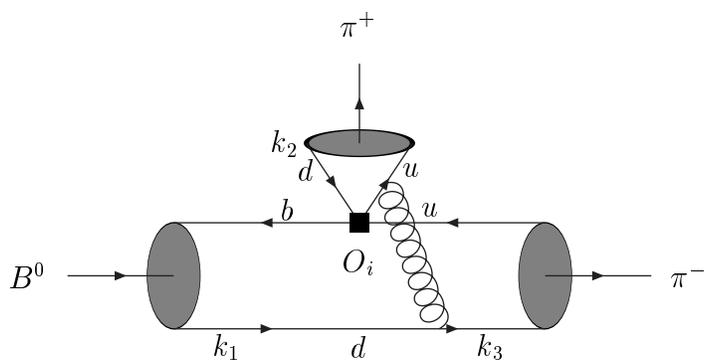
\item Fig. 2: Lowest-order nonfactorizable diagrams for
the $B_d^0 \to \pi^{+}\pi^{-}$ decay.

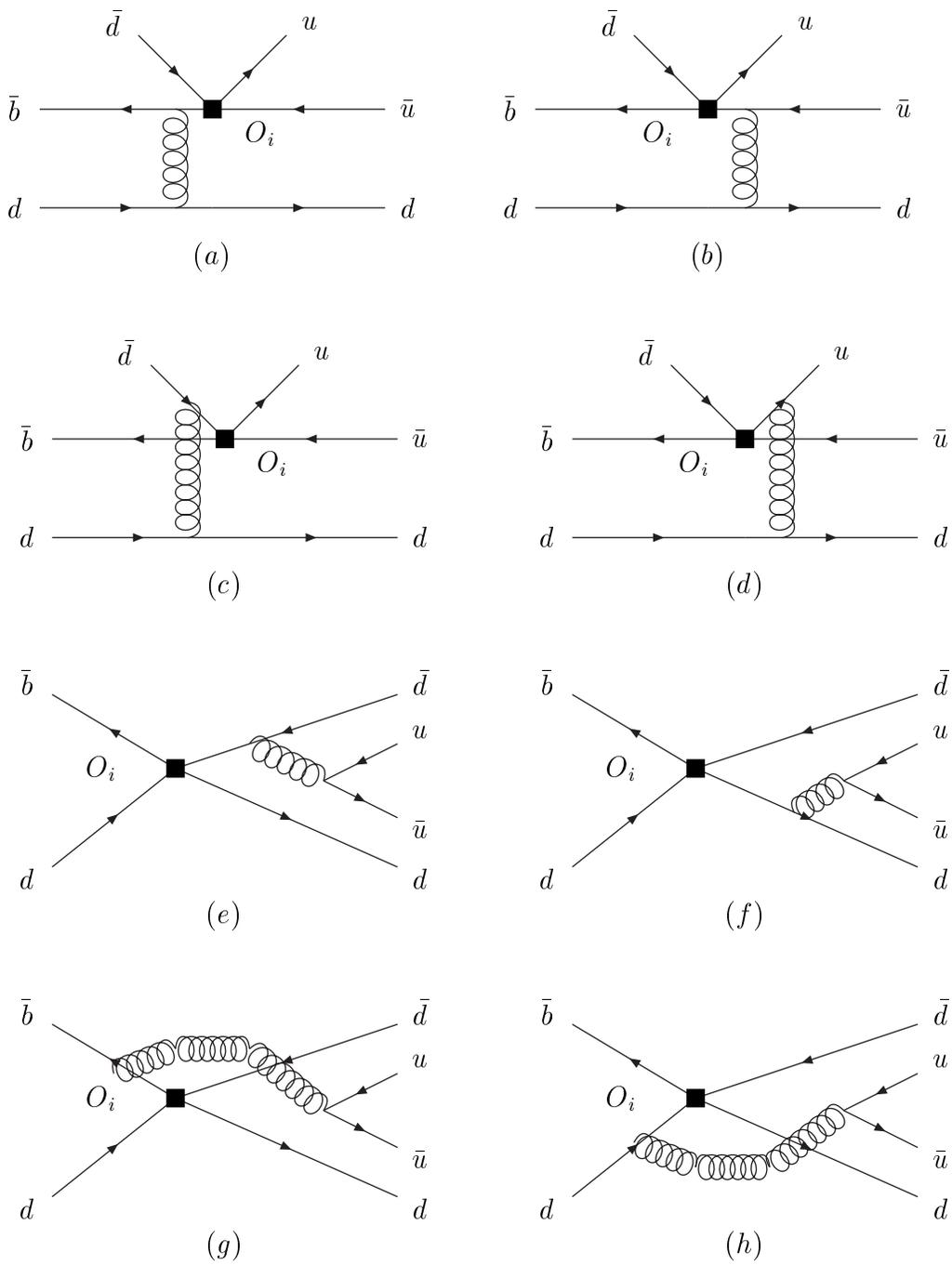
\item Fig. 3: Lowest-order hard amplitudes for 
the $B_d^0 \to \pi^{+}\pi^{-}$ decay.

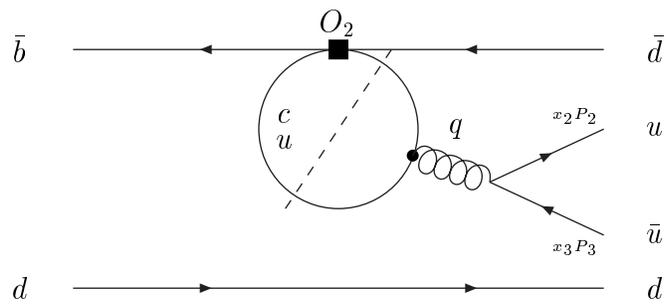
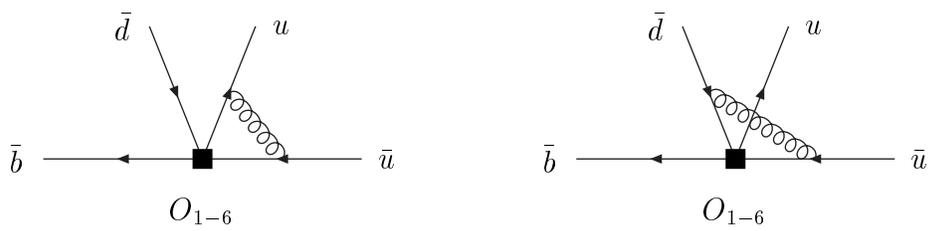
\item Fig. 4: Diagrams that give strong phases in the FA and
BBNS approaches.

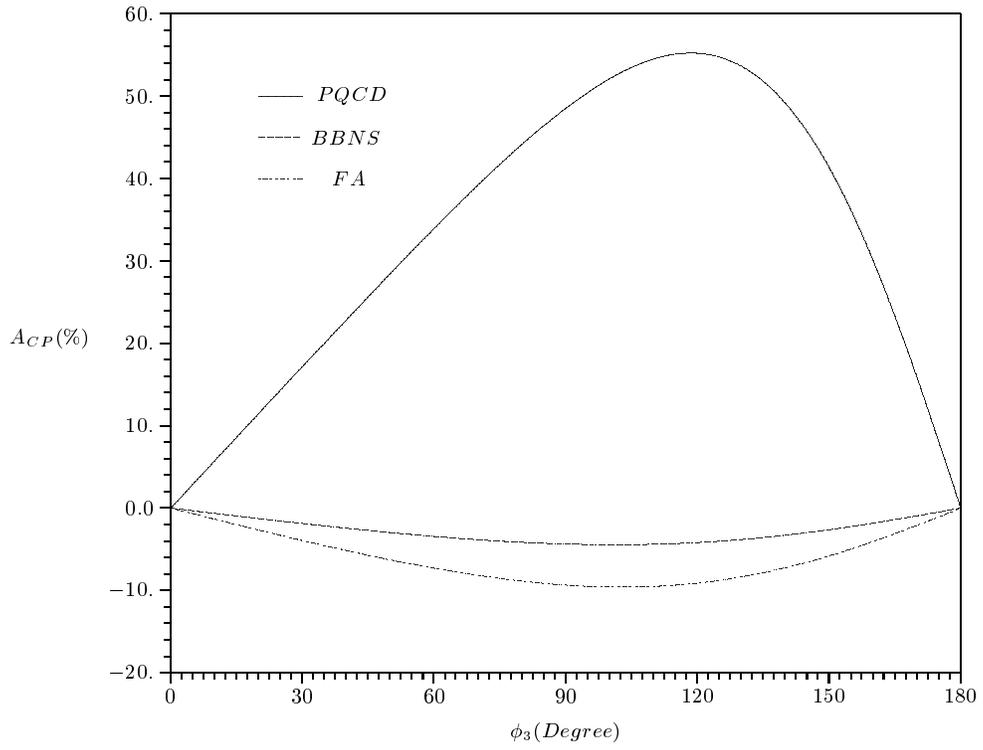
\item Fig. 5: Direct CP asymmetries in 
the $B_d^0 \to \pi^{\pm}\pi^{\mp}$ decays.

\end{enumerate}

\newpage

\begin{table}
\begin{tabular}{|c|ccc|c|}
Amplitudes & Left-handed column & 
Right-handed column & PQCD & BBNS \\
\tableline 
$Re(f_{\pi} F^T)$ & (a)\hspace{0.5cm}$4.39 \cdot 10^{-2}$ & 
(b)\hspace{0.5cm}$2.95 \cdot 10^{-2}$
 & $7.34 \cdot 10^{-2}$ &  $7.57 \cdot 10^{-2}$    \\
$Im(f_{\pi} F^T)$ & $-$  & $-$ 
 & $-$ &  $ 1.13 \cdot 10^{-3}$    \\
\tableline
$Re(f_{\pi} F^P)$ & (a) \hspace{0.5cm}-$3.54 \cdot 10^{-3}$ & 
(b)\hspace{0.5cm}-$2.33 \cdot 10^{-3}$ 
& -$5.87 \cdot 10^{-3}$ & -$3.04 \cdot 10^{-3}$ \\
$Im(f_{\pi} F^P)$ & $-$ & $-$ 
& $-$ & -$1.27 \cdot 10^{-3}$ \\
\tableline
$Re(f_{B} F_a^P)$ & (e)\hspace{0.5cm}$5.05 \cdot 10^{-4}$ & 
(f)\hspace{0.5cm}-$1.94 \cdot 10^{-3}$
& -$1.42 \cdot 10^{-3}$ & $-$  \\
$Im(f_{B} F_a^P)$ & (e)\hspace{0.5cm}$2.19 \cdot 10^{-3}$ & 
(f)\hspace{0.5cm}$3.72 \cdot 10^{-3}$ 
& $5.91 \cdot 10^{-3}$ & $-$ \\
\tableline
$Re(M^T)$ & (c)\hspace{0.5cm}$5.02 \cdot 10^{-3}$ & 
(d)\hspace{0.5cm}-$6.55 \cdot 10^{-3}$ 
& -$1.53 \cdot 10^{-3}$ & -$7.71 \cdot 10^{-4}$ \\
$Im(M^T)$ & (c)\hspace{0.5cm}-$3.83 \cdot 10^{-3}$ & 
(d)\hspace{0.5cm}$7.03 \cdot 10^{-3}$
& $3.20 \cdot 10^{-3}$ & $-$ \\
\tableline
$Re(M^P)$ & (c)\hspace{0.5cm}-$2.29 \cdot 10^{-4}$ & 
(d)\hspace{0.5cm}$2.75 \cdot 10^{-4}$
& $4.66 \cdot 10^{-5}$ & $4.17 \cdot 10^{-3}$ \\
$Im(M^P)$ & (c)\hspace{0.5cm}$1.95 \cdot 10^{-4}$ & 
(d)\hspace{0.5cm}-$3.08 \cdot 10^{-4}$
& -$1.13 \cdot 10^{-3}$ & $-$ \\
\tableline
$Re(M_a^P)$ & (g)\hspace{0.5cm}$1.14 \cdot 10^{-5}$ & 
(h)\hspace{0.5cm}-$1.48 \cdot 10^{-4}$
& -$1.37 \cdot 10^{-4}$ & $-$ \\
$Im(M_a^P)$ & (g)\hspace{0.5cm}-$9.12 \cdot 10^{-6}$ & 
(h)\hspace{0.5cm}-$1.27 \cdot 10^{-4}$ 
& -$1.36 \cdot 10^{-4}$ & $-$ \\
\end{tabular}
\label{TABLE11.1}
\caption{Amplitudes for the $B_d^0 \to \pi^{+} \pi^{-}$ decay from 
Fig.~3, where $F$ ($M$) denotes factorizable (nonfactorizable) 
contributions, $P$ ($T$) denotes the penguin (tree) contributions,
and $a$ denotes the annihilation contributions. Here we adopted 
$\phi_3=90^0$, $R_b=0.38$, and $\alpha_s(m_b)= 0.2552$
in the numerical analysis for the BBNS approach.  }
\end{table}

\end{document}